# Security Issues in Controller Area Networks in Automobiles

Robert Buttigieg, Mario Farrugia and Clyde Meli
University of Malta
Msida Malta
mario.a.farrugia@um.edu.mt

*Abstract*—**Modern vehicles may contain a considerable number of ECUs (Electronic Control Units) which are connected through various means of communication, with the CAN (Controller Area Network) protocol being the most widely used. However, several vulnerabilities such as the lack of authentication and the lack of data encryption have been pointed out by several authors, which ultimately render vehicles unsafe to their users and surroundings. Moreover, the lack of security in modern automobiles has been studied and analyzed by other researchers as well as several reports about modern car hacking have (already) been published. The contribution of this work aimed to analyze and test the level of security and how resilient is the CAN protocol by taking a BMW E90 (3-series) instrument cluster as a sample for a proof of concept study. This investigation was carried out by building and developing a rogue device using cheap commercially available components while being connected to the same CAN-Bus as a man in the middle device in order to send spoofed messages to the instrument cluster**.

## I. INTRODUCTION

Modern automobiles contain several ECU's which are connected together using a network protocol. Controller Area Network (CAN), Local Interconnect Network (LIN), Media Oriented Systems Transport (MOST) and FlexRay are all types of in-vehicle networks, with CAN being the most used nowadays [1] – [3]. An ECU is a device which controls one or more aspects of the electrical system inside a vehicle, such as the engine, brakes, airbags and lights.

Manufacturers started replacing mechanical parts with electronic components due to strict rules with regards to emissions for efficiency purposes and improving engine performance [4] and [5]. Consequently, vehicles ended up having several devices connected to each other using point-to-point connections, which resulted in an increase in complexity and the number of wires used. This problem was solved by implementing the CAN protocol since a robust in-vehicle network was required [5]. Furthermore, due to the increase in the information transfer between the connected devices, vehicles often have multiple networks which connect different subsets of ECU's [1]. Data provided over the vehicle network can be very useful in diagnosis as well as research [6].

One issue with CAN is that security was not even considered during its design stages since it was thought that vehicles would remain closed systems [7] and [8]. When vehicles were purely mechanical, the only way how one could gain access to a car is by physical access. Unfortunately, this is not the case in today's modern vehicles. External attack surfaces of modern automobiles allow the possibility of a remote exploitation through various means of channels such as diagnostic tools, CD player, Bluetooth and cellular radio [9]. Considering that current automobiles rely heavily on software due to the introduction of various ECU's, there are high chances that the code may contain bugs or vulnerabilities which when exploited, they can lead to potential risks, thus rendering vehicles unsafe. Moreover, due to lack of authentication mechanisms in the CAN protocol, an attacker may also connect an external malicious device directly with the CAN-Bus with the intention to harm passengers or even cause dangerous accidents.

## II. CONTROLLER AREA NETWORK

CAN is a robust serial communications bus which was designed by Bosch in the mid-1980 and has a maximum bitrate of one megabit per second [1] and [10] – [13]. One of the main goals why CAN was invented is to reduce the wiring costs and complexity inside the vehicle [1] and [10] – [12]. It is most suitable for systems where a small amount of information needs to be exchanged [1], [11] and [12]. Although it was originally intended for the automotive industry, it is also being used in other industries and control systems, such as medical devices, elevators, robotics, building automation and manufacturing [1], [4], [11] and [12]. The CAN protocol was subsequently adopted as an ISO standard (11898) in 1993 [1] and [4].

CAN is implemented as a pair of differentially balanced signalled wires (the electrical current in both wires is equal but opposite in direction), CAN_H and CAN_L [1], [4] and [12]. Having a balanced differential signalling results in a field cancelling effect, providing high noise immunity, fault tolerance and reduces electromagnetic interference [1] and [12]. Bits transmitted over CAN-Bus can be either dominant (logical zero) or recessive (logical one), where dominant bits win arbitration over recessive bits [1], [4] and [10] – [12]. During recessive states, CAN_H and CAN_L are both 2.5V while in dominant states, CAN_H goes up to 5V whereas CAN_L goes to ground [4].

The CAN protocol enables prioritisation of messages based on their identifiers [1], [4] and [10] – [12], where the lower the identifier, the higher the priority [1], [4], [11] and [12]. A CAN message can transmit up to eight bytes of data [1], [4], [10] and [12]. Moreover, CAN is a message-based network rather than address-based and therefore, it does not provide any way how a node can transmit a message to only a specific node [1], [11] and [12]. Instead, it is a broadcast type of network, meaning that each message is broadcasted to all other nodes. Every node is built with a message filtering mechanism which filters messages based on their identifiers and considers only those which are relevant to it while ignoring others. Furthermore, nodes on the network can be added or removed easily since no modifications in the hardware, software or the application layer are required [1], [10] and [11].

CAN is a Carrier Sense Multiple Access with Collision Detection (CSMA/CD) protocol [1], [4], [11] and [12]. When a node wants to send messages over the network, it must monitor the serial bus first, and if the bus is idle, it can start transmitting. If two nodes start transmitting at the same time, a collision is detected. This issue is solved by a non-destructive bitwise arbitration, where each transmitter compares its transmitted bit with the bit that is monitored on the bus. The node which has lost arbitration has to abort transmission. As a result, messages with the highest priority are transmitted first, while the other messages with a lower priority have to wait until the bus is idle again [1], [4] and [10] – [12]. Furthermore, the CAN protocol supports four different message types; overload, error, remote and data frames, with the latter being the mostly used since it is used to transmit messages across the network. Data frames can be categorized in standard and extended formats, where the standard frame consists of an 11-bit identifier while the extended format contains a 29-bit identifier [1] and [10] – [12].

III. SECURITY ISSUES IN THE CAN PROTOCOL

While CAN offers several mechanisms for error detection, data integrity and data consistency, it does not provide any form of secure communication across the network. The computerization of automobiles has brought several advantages such as driver comfort, vehicle efficiency and performance. Despite these benefits, this dependence on such devices have broadened potential attack surfaces and exposed many vulnerabilities as well [5]. Until now, most manufacturers have adopted the approach of 'security by obscurity', meaning that the implementation is kept private so as no one can fully understand it and possibly manipulate it [5] and [13]. However, it does no longer apply in today's automotive world since certain vulnerabilities may be well-known by attackers. Several authors agree that the CAN protocol lacks security and it does not provide any secure methods against malicious attacks [2], [3], [5], [7], [8], [13] and [14]. Others argue that while CAN was designed for reliability, it was not even designed with security in mind [3], [4], [7], [8] and [13]. Furthermore, according to Hiroshi et al. [2], message spoofing is considered as one of the main threats to in-vehicle networks since it is possible to display a falsified value to a vehicle's speedometer or tachometer or even taking control of critical safety systems.

*A. Lack of Authentication*

CAN does not provide any message or device authentication in order to identify the sender of a particular message or to determine the identity of an ECU [2] – [5], [7], [8], [13] and [14]. This means that the CAN protocol is not capable of distinguishing between a legitimate ECU and a malicious one [7] and [14]. This security weakness makes replay attacks and transmission of spoofed messages possible since an unauthorized device can be easily connected to the CAN-Bus [2], [4], [5], [7], [13] and [14].

*B. Lack of network segmentation*

Originally, CAN was invented as a single serial bus connecting all nodes within a single network. Since all messages are broadcasted over the network, any ECU can send messages to other critical safety ECU's, such as the braking control unit [13]. This vulnerability makes it easier for unauthorized devices to also communicate with other ECU's which are responsible for the safety of the vehicle. Additionally, if an attacker gains access to the CAN cables, s/he may have access to the whole network [5].

*C. Lack of data encryption*

Another significant vulnerability in the CAN protocol is the lack of data encryption. CAN does not provide any encryption mechanisms to safeguard the integrity and privacy of the data [5]. Due to the limited computational power of ECU's, it is difficult to implement robust cryptographic algorithms [3]. Since CAN does not either provide any form of authentication, it makes it even possible for an authorized device to listen to the network traffic [5].

*D. Vulnerable to Denial of Service Attacks*

CAN is also vulnerable to denial of service attacks. This attack can be performed by transmitting consecutive high priority messages by sending successive dominant bits on the bus [3], [4], [8] and [14]. Such attacks can result in flooding the bus, bringing the network down or occupying all bandwidth which will make other nodes unable to transmit any messages [8] and [14].

IV. EXPLOITING CAN VULNERABILITIES

Due to the lack of authentication in the CAN protocol, it is possible to masquerade an ECU or replace a legitimate ECU with a malicious one using a hardware device [8]. A device such as a hardware circuit board can also be attached to the CAN-Bus since the network wires of most vehicles are exposed in easy to find locations [14]. Hiroshi et al. [2] describe two attack scenarios how spoofing messages on the CAN-Bus can be performed. The first scenario illustrates an attack where the original program of an ECU is replaced by a malicious one while the other scenario represents an attack where an unauthorized device is connected to the CAN-Bus. Both scenarios are depicted in Fig. 1 and Fig. 2.

The European Union Agency for Network and Information Security (ENISA) is responsible for the security in networks and information in the EU and aims to assist and advocate good practices with regards to security and resilience of critical systems. Lately, a report was published by this agency to identify cybersecurity and the resilience of smart vehicles. The report describes the lack of security in current in-vehicle networks and presents the reader with a series of possible threats and attack models which can be performed on the network. Amongst others, the report mentions that a man in the middle attack is possible by connecting an unauthorized device directly with CAN-Bus. Moreover, the report also states that replay attacks allow the attacker to identify commands which control critical safety systems and also indicates the vulnerability to denial of service attacks [15].

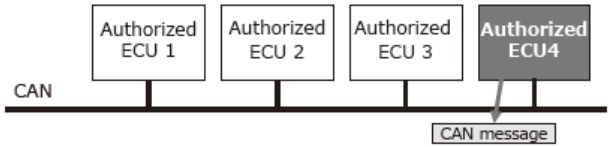

Fig. 1. Substitution of an authorized ECU program.

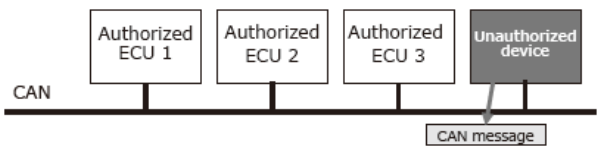

Fig. 2. Connection of an unauthorized device.

Modern vehicles provide a diagnostic port, called the OBD-II (On-Board Diagnostic) port which is located under a vehicle's dashboard. It allows technicians to perform diagnosis on the in-vehicle network, testing of emissions control and report any faults [3] – [5], [8] and [6]. The implementation of this port in automobiles has become obligatory both in the US (from 1996) and the EU (since 2001) [3], [4] and [15]. Since no specifications are provided with regards to the implementation of this port, some manufacturers such as the BMW dedicate a different sub-network bus (called the Diag bus), while others implement it in a way that it is directly connected to the CAN-Bus. However, the latter broadens the vehicle attack surface since it guarantees direct access to the network [4], [7], [8] and [15]. Due to this exposed potential threat, security researchers have demonstrated actual attacks to highlight its vulnerability. Although OBD was not the main focus of this project, it shows what an external device is capable of doing when provided with access to the CAN-Bus.

Charlie Miller (director of vehicle security research for IOActive) and Chris Valasek (security researcher for Twitter) were instructed to research automobile security where they managed to inject spoofed messages onto the CAN-Bus of a Toyota Prius (2010) and Ford Escape (2010) [5]. This attack was possible since the CAN protocol does not provide any form of authentication and messages are broadcasted onto the whole network. This experiment is very similar to what Koscher *et al.* have proved. However, the published report includes a detailed description of how the attack was carried out and upon which cars the attack was tested. The report also includes the CAN architecture implemented in both cars and their respective wiring diagrams, the code used to perform the attack, and the CAN message IDs of both cars together with their respective functions [16].

The setup of this attack consisted of a laptop connected to the OBD-II port. Once they managed to read and write to the CAN-Bus, message ID's were identified by performing a replay attack. The results of this experiment concluded that it is possible to inject spoofed messages to critical safety ECU's for both vehicles. Amongst others, this resulted in disabling or applying the brakes, killing the engine, display forged values to the instrument cluster, locking and unlocking the doors and tampering with both interior and exterior lights [16].

## V. AIMS AND OBJECTIVES

The main objective of this paper was to analyze the security in controller area networks in modern cars. This was done by connecting a prototype rogue device to the CAN-Bus as a man in the middle attack and investigate its resilience to malicious attacks. As a proof of concept and for demonstration purposes only, a BMW E90 (3-series) instrument cluster was used. A vehicle simulator was implemented in order to be able to simulate the rest of the car. The simulator was made to work both in a demo mode and in a manual mode. The latter consists of a hardware interface consisting of various switches which control all gauges and other critical systems of the instrument cluster. On the other hand, the role of the rogue device was to modify the data which was being transmitted in order to test whether the instrument cluster provides any form of in-built security.

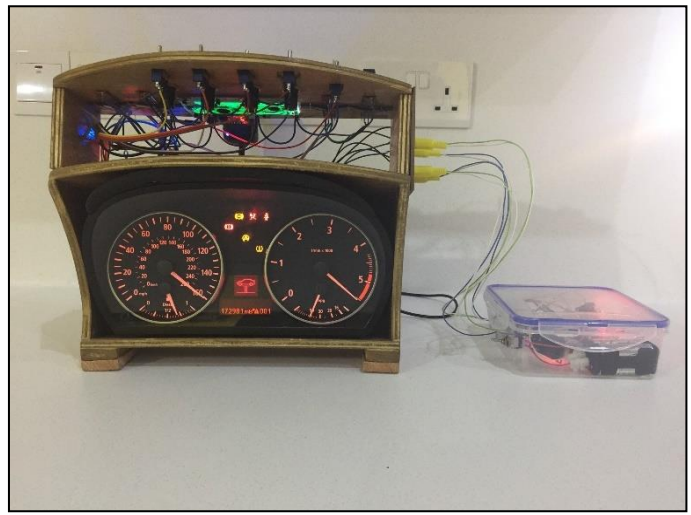

Fig. 3. Photo of instrument cluster being hacked by a rogue device in a transparent plastic box, displaying an rpm of 5500, 260km/hr and both airbag and ABS system disabled.

## VI. DESIGN

The main components of the design of the project are the vehicle simulator, the BMW E90 instrument cluster and the rogue device, which are all connected using the CAN_H and CAN_L wires, as shown in Fig. 4. The reason why the BMW group was chosen for this proof of concept is that it is a reputable brand, which strives to implement secure elements in its products. The instrument cluster was chosen amongst other ECU's inside the vehicle because it is one of the main ECU's which the driver interacts with and it displays information related to the car in an understandable way, even for non-technical people. The vehicle simulator is responsible for the simulation of the rest of the car by sending the appropriate and relevant CAN messages over the network as if the cluster was installed in a real car. Meanwhile, the rogue device is implemented as a man in the middle device between the vehicle simulator and the instrument cluster. Its function is to receive incoming packets from the vehicle simulator, change the data which are being transmitted and send them to the instrument cluster. The instrument cluster is powered using an AC/DC adapter while the rogue device will be using a battery power supply.

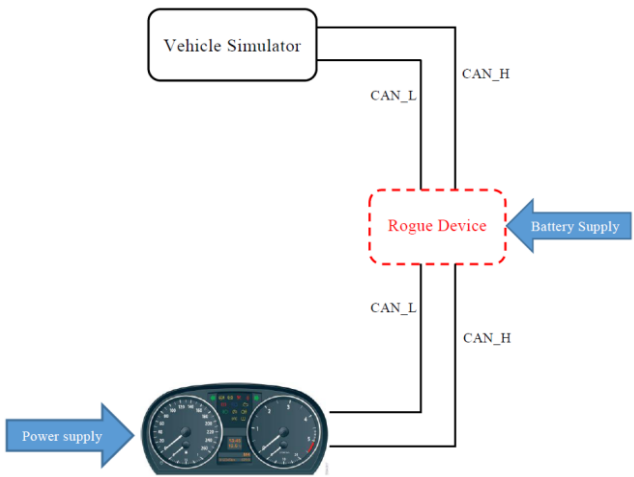

Fig. 4. A block diagram of the research project design.

## VII. IMPLEMENTATION

The implementation stage is categorized into three main parts. The first part describes the reasons why and how the instrument cluster ECU was chosen while the second and third parts describe in detail the vehicle simulator and the rogue device.

### A. BMW E90 (3 Series) Instrument Cluster

The BMW E90 (3 Series) car was chosen since it is quite a common car and has a good reputation of being well engineered and solidly built. After getting familiar with the CAN architecture implemented by the BMW and analyzing wiring and system circuit diagrams, the instrument cluster ECU was chosen as a bench test for this experiment since it is not dependent on too many other ECU's, therefore making it easier to be tested as a standalone. The BMW group has already implemented a security feature where a gateway module is placed between the different bus systems and several message counter mechanisms to make replay and man in the middle attacks more difficult. These message counters work in such a way that if the counter of a particular CAN message is not being incremented properly, an error is generated.

### B. Vehicle Simulator

The setup for the vehicle simulator consisted of a CAN-Bus shield attached to an Arduino MEGA 2560 board, both connected to the instrument cluster using an 18-pin ribbon cable. The power supply and ground wires are both connected to the Arduino board while the CAN_H and CAN_L wires are connected with the CAN-Bus shield. Power was supplied to the Arduino using an AC/DC adapter with an output of 12V 2A.

Messages were sent onto the CAN-Bus using the `sendMsgBuf(INT32U id, INT8U ext, INT8U rtr, INT8U len, INT8U *buf)` method from the CAN-Bus library. The code was structured in a way that every CAN ID has its own method where the data to be transmitted is stored in a byte array. The data inside every CAN message change according to the current status of the car such as engine running or engine off. This helps to eliminate any hard-coded messages as well as making it easier to build a user interface for the simulator. Furthermore, in order to send messages according to their specific timestamp, each CAN ID message has its own variable which is responsible for handling message timings.

Due to the lack of information with regards to how the instrument cluster can be controlled and more significantly the unknown message timestamps, research was conducted to try to find a CAN log from the same model car with correct message timings. Fortunately, a CAN log from the K-CAN of an E90 BMW car was acquired. This log consists of both the CAN message ID and the data transmitted, but no information regarding timings is provided. The only option was to decode and analyze this log in order to determine message timestamps by using the 0x130 message (which is sent every 100 milliseconds) both as a starting point and a reference point. To facilitate the identification of the messages required as well as the timings at which messages have to be sent, a program using C# was developed.

The BMWSimulator program resides on the Arduino board and is responsible for the simulation of the car by sending the appropriate CAN messages according to their respective timings to the instrument cluster. The program initialises the CAN-Bus at 100Kbits/s and all message counters and message timer variables to their appropriate value. Table 1 describes the CAN message ID's which are sent to the instrument cluster together with their data length, description and their respective timestamps. Furthermore, the vehicle simulator can work both in a demo mode and in a manual mode, where a toggle switch is used to alternate between the two modes. The demo mode shows how the instrument cluster works in a real-life scenario while the engine is running. On the other hand, the two main aims of implementing the manual mode are for demonstration purposes and to show that several aspects of the instrument cluster can be controlled independently, irrespective whether it is possible or not in real life. The manual mode consists of a hardware interface made up of various toggle switches, potentiometers and push button switches which control all gauges and other critical systems of the instrument cluster. A display screen was also used to show the current status of the car.

Table 1: Description of the CAN ID's, data length, description and the timestamp at which each message has to be sent.

| CAN ID (HEX) | Length (Bytes) | Description | Timestamp (Milliseconds) |
|---|---|---|---|
| 0A8 | 8 | Torque, Clutch and Brake Status | 10 |
| 0AA | 8 | Engine RPM and Throttle Position | 10 |
| 0C0 | 2 | ABS / Brake Counter | 200 |
| 0CE | 8 | Individual Wheel Speeds | 10 |
| 0D7 | 2 | Counter (Airbag / Seatbelt related) | 200 |
| 130 | 5 | Ignition and Key Status (Terminal 15) | 100 |
| 19E | 8 | ABS / Braking Force | 200 |
| 1A6 | 8 | Speed | 100 |
| 1D0 | 8 | Engine Temperature, Pressure Sensor and Handbrake | 200 |
| 21A | 3 | Lighting Status | 5000 |
| 26E | 8 | Ignition Status | 200 |
| 335 | 8 | Unknown | 1000 |
| 349 | 5 | Fuel Level Sensors | 200 |
| 34F | 2 | Handbrake Status | 1000 |
| 380 | 7 | VIN Number | Once |
| 39E | 8 | Set Time and Date | Once |
| 3B4 | 8 | Battery Voltage and Charge Status | 4000 |
| 581 | 8 | Seatbelt Status | 5000 |

*C. Rogue Device*

The scope of the rogue device was to be implemented as a man in the middle, between the vehicle simulator and the instrument cluster in order to investigate the level of security implemented in the CAN protocol. This device received incoming packets from the vehicle simulator, modify certain messages and send them to the instrument cluster. This experiment was carried out to evaluate whether the instrument cluster ECU detects an unauthorized device on the CAN-Bus and whether it prevents the device from displaying forged values to the tachometer and speedometer and disable both the ABS and the airbags. Moreover, power was supplied to the rogue device using a battery in order to demonstrate the ease with which such a device could be implemented. This device can be remotely controlled using any chosen means such as RF, GSM, or Wi-Fi.

These two CAN-Bus shields were connected to the NodeMCU by referring to the pinout diagram and connecting the SPI pins, CS pins (one for each CAN-Bus shields) and the positive and negative with their respective pins. Since the NodeMCU and the CAN-Bus shields have different clock speeds, the original CNF register values from the CAN-Bus library had to be modified with other values.

Since the NodeMCU provides a Wi-Fi on-board connection, an interface for the hacker was developed which enables the attacker to establish a connection with the rogue device to remotely send forged messages to the instrument cluster. The BMWHacker program resides in the NodeMCU board and it provided the attacker with a UDP connection to the BMW Hacker interface. The BMW Hacker interface is a program, developed using C#, which enables the attacker to monitor the status of the car in real time and to select a forged value for the speedometer and tachometer and with the option of disabling the airbags and the ABS. These values are then sent to the rogue device which will modify the CAN messages in order to perform the attack.

The instrument cluster has been successfully hacked by modifying the data within the CAN message in order to display falsified values to the speedometer and tachometer. Moreover, the airbag system and the ABS were also hacked by blocking certain messages from being transmitted. The latter is possible because when the instrument cluster stops receiving certain messages, an error is displayed.

Fig. 5. Screenshot of the BMW hacker program.

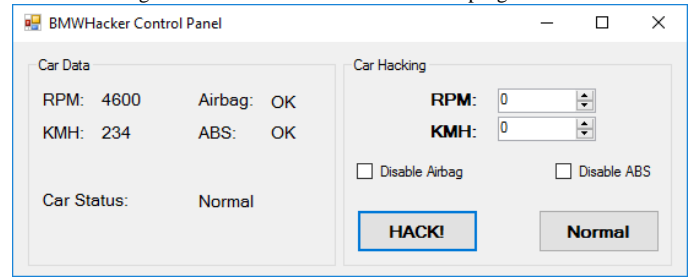

VIII. RESULTS AND EVALUATION

The main aim of this research was to look into the lack of security in the CAN-Bus protocol, in particular, the lack of authentication of devices and the lack of data encryption. This experiment focused on the possibility that an unauthorized man in the middle device can be connected to the CAN-Bus with the ability to send spoofed messages. As a result, these weaknesses can give the attacker access to the whole network and the ability to write and read CAN messages.

Results have shown that the instrument cluster did not detect the unauthorized device, which was successfully connected to the CAN-Bus by simply receiving incoming messages and sending them over the network. The instrument cluster neither detected that messages were being sent from an illegitimate device since it behaved in the same way when messages were directly sent from the vehicle simulator. Furthermore, the instrument cluster was not able to detect scenarios which are not possible in real life, such as switching on the main beam lights without the side lights and driving at maximum speed with no fuel. On the other hand, the instrument cluster managed to detect that the handbrake was on while driving.

## IX. CONCLUSION

Even though BMW strives to implement secure in-vehicle networks, it is still vulnerable to attacks. The use of message counters implemented by BMW and the fact that the diagnostic port is connected to a separate bus makes it difficult to perform such attack using the OBD-II port. However, the contribution of this work shows that it is still possible by simply connecting an external device directly to the CAN-Bus.

Despite the fact that this demonstration was performed on a BMW E90 instrument cluster, it implies that such attacks can be performed on other car manufacturer's products as well, as long as they use the CAN-Bus protocol in their in-vehicle network. Reputable car manufacturers have been implementing certain mechanisms to make reverse engineering and attacks more difficult, such as the use of message counters in the case of the BMW group. Even though this study was aimed to create awareness regarding the lack of security in the CAN-Bus protocol, with enough knowledge, time and resources, an attacker can still perform such attacks with the intentions of harming passengers and cause dangerous road accidents.

Although manufacturers have acknowledged the lack of security in in-vehicle networks, further research is still required to provide a more secure automotive environment and to prevent malicious attacks. This project can be further continued to prove that denial of service attacks are practically possible in the CAN-Bus protocol as well as showing their consequences. Furthermore, research could also analyze the possibility of implementing a security mechanism in order to be able to withstand both man in the middle and denial of service attacks. This security feature may take into consideration both private and public encryption keys which will be encoded in the firmware of all ECU's, therefore making it more difficult for replay attacks and for attackers to send or read CAN messages over the network.